\documentclass[11pt,a4paper,epsf,epsfig,psfrag]{article}
\usepackage{jheppub}
\usepackage{amsmath,graphicx,amsfonts, mathrsfs,amssymb}
\usepackage{amsmath,epsfig}
\usepackage{amssymb,amsfonts}
\usepackage{color}
\usepackage{latexsym}
\usepackage{epsfig}
\usepackage{pdfsync}
\usepackage{marvosym}
\newbox\pippobox

\title{Pion condensation in a soft-wall AdS/QCD model}

\author[a,b]{Meng Lv}
\author[c,1]{Danning Li\note{Corresponding author}}
\author[d,e]{Song He}

\affiliation[a]{College of Physical Science and Technology, Sichuan University, Chengdu 610064, P.R. China}
\affiliation[b]{Key Laboratory of Radiation Physics and Technology of Ministry of Education, Sichuan University, Chengdu 610064, P.R. China }
\affiliation[c]{Department of Physics and Siyuan Laboratory, Jinan University, Guangzhou 510632, P.R. China}
\affiliation[d]{Max Planck Institute for Gravitational Physics (Albert Einstein Institute) Am M\"{u}hlenberg 1, 14476 Golm, Germany}
\affiliation[e]{Center for Theoretical Physics and College of Physics, Jilin University, Changchun 130012, P. R. China}

\emailAdd{lvmengphys@scu.edu.cn}
\emailAdd{lidanning@jnu.edu.cn}
\emailAdd{hesong17@gmail.com}

\abstract{Finite isospin chemical potential $\mu_I$ and temperature $T$ have been introduced in the framework of  soft-wall AdS/QCD model. By self-consistently solve the equation of motion, we obtain the phase boundary of pion condensation phase, across which the system undergoes a phase transition between pion condensation phase and normal phase. Comparing the free energy of solutions with and without pion condensation, we find that the phase transition is of first order type both at large $\mu_I$ and small $\mu_I$. Qualitatively, the behavior at large $\mu_I$ is in agreement with the lattice simulation in Phys.Rev.D66(2002)034505, while the behavior at small $\mu_I$ is different from lattice simulations and previous studies in hard wall AdS/QCD model. This indicates that  a full back-reaction model including the interaction of gluo-dynamics and chiral dynamics might be necessary to describe the small $\mu_I$ pion condensation phase.  This study could provide certain clues to build a more realistic holographic model. }
\keywords{Finite isospin chemical potential, chiral condensation, pion condensation, soft-wall AdS/QCD}

\begin{document}
\maketitle
\section{Introduction}
\label{sec-int}

Phase transitions of Quantum Chromodynamics(QCD) with finite temperature and isospin density $n_I=n_u-n_d$, which describes the asymmetry between up and down quarks as well as protons and neutrons, have attracted many attentions in the past decades. The isospin asymmetry exists widely in many different situations.  For example, the proton-to-neutron ratio in relativistic heavy ion collisions(RHIC) is about $2/3$ in Au or Pb beams, which might causes the imbalance between the generated charged pions\cite{Li:1997px}. In astronomy, the formation of neutron stars is strongly related to the equation of state at finite isospin chemical potential\cite{Lattimer:2006xb,Steiner:2004fi}. In the early Universe, the isospin asymmetry in baryon sector coming from the lepton asymmetry could affect the equilibrium conditions around the time of QCD transition\cite{Schwarz:2009ii}.

\begin{figure}[h]
\begin{center}
\epsfxsize=6.5 cm \epsfysize=6.5 cm \epsfbox{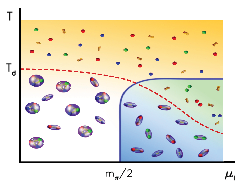}
\hspace*{0.1cm} \epsfxsize=6.5 cm \epsfysize=6.5 cm \epsfbox{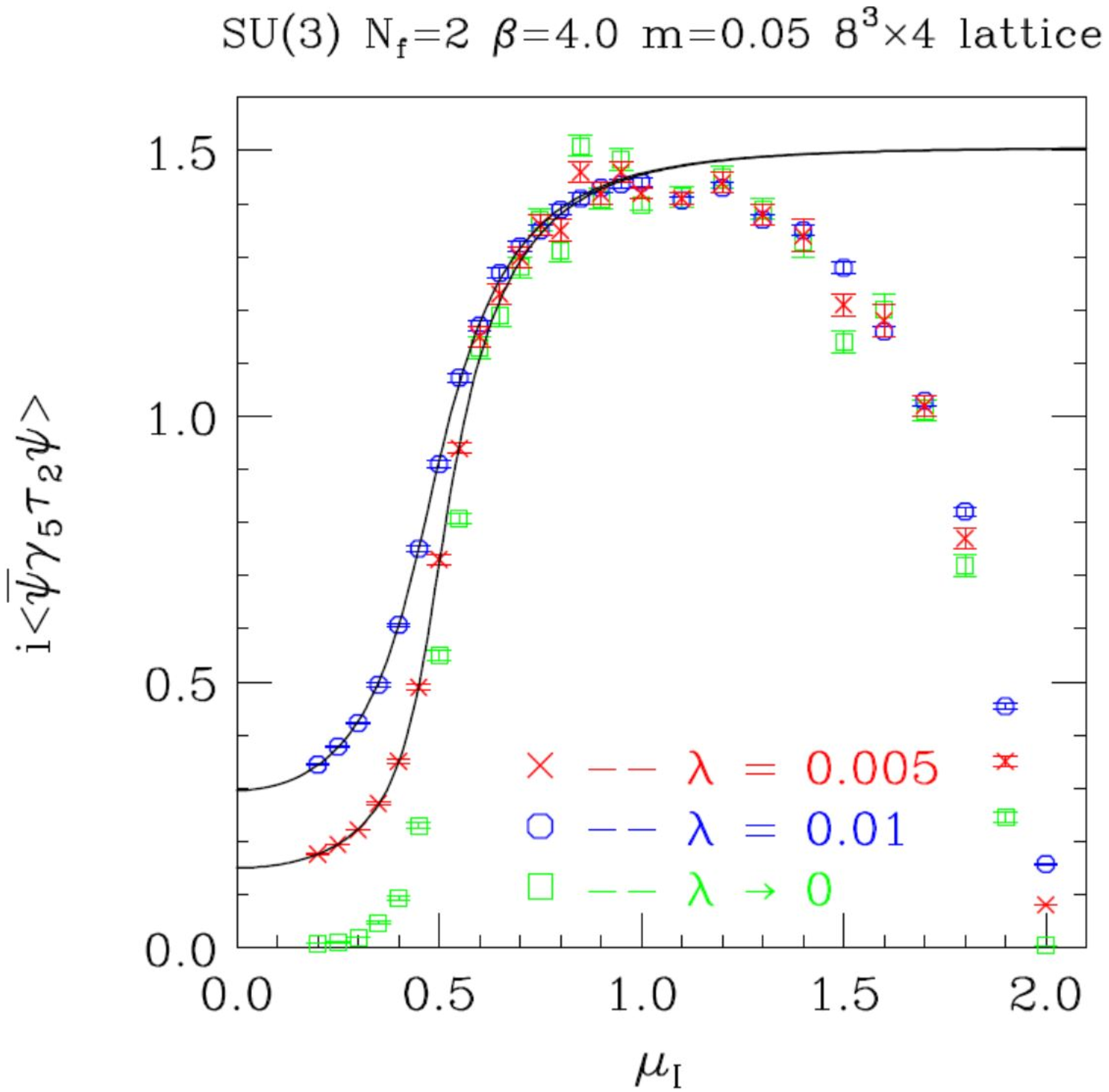}
\vskip -0.05cm \hskip 0.15 cm
\textbf{( a ) } \hskip 6.5 cm \textbf{( b )} \\
\end{center}
\caption{Sketch plot of phase diagram in $\mu_I-T$ plane(Panel.(a), taken from \cite{Brandt:2017oyy}) and pion condensation as functions of isospin chemical potential $\mu_I$ at zero temperature from lattice simulation(Panel.(b), taken from \cite{Kogut:2002zg}).  }
\label{lattice-phase}
\end{figure}

The non-perturbative physics dominant near transition point requires methods beyond traditional perturbative expansion. Many efforts have been made in this area, including lattice simulations(LQCD)\cite{Alford:1998sd,Kogut:2002zg,Detmold:2012wc,Cea:2012ev,Brandt:2017oyy}, functional renormalization group(FRG)\cite{Svanes:2010we,Wang:2015bky}, chiral perturbation theory($\chi$PT)\cite{Son:2000xc,Kogut:2001id}, perturbative methods\cite{Graf:2015pyl}, random matrix models\cite{Klein:2003fy,Klein:2004hv}, effective models\cite{Toublan:2003tt,Barducci:2004tt,Xia:2013caa,He:2005nk,Carignano:2016lxe,Kamikado:2012bt,Phat:2011zza} and so on.  Different from baryon number density, finite isospin chemical potential would not cause sign problem, due to the opposite signs of the chemical potentials for $u$ and $d$. Therefore, lattice simulations could be safely applied in this region. Currently, the lattice simulations seem to support the $\chi$PT study, which predicts a second order phase transition from normal phase to pion condensation phase at zero temperature. The critical value of $\mu_I$ is around half of pion mass $m_\pi$. However, different from monotonous rise of $\mu_{I,c}(T)$ with the increasing of $T$ as predicted by $\chi$PT, $\mu_{I,c}(T)$ predicted by lattice simulations has a flatten pattern at the intermediate region of  $T$ \cite{Brandt:2017oyy}, as shown in Fig.\ref{lattice-phase}(a). Moreover, at zero temperature, pion condensation does not grow with $\mu_I$ monotonously. The growth will ceased at ceratin $\mu_I$ and pion condensation starts to decrease to zero\cite{Kogut:2002zg} at large $\mu_I$(see Fig.\ref{lattice-phase}(b), taken from \cite{Kogut:2002zg}). Also, the simulations in \cite{Kogut:2002zg} support a first order phase transition at large $\mu_I$.

%\begin{figure}[h]
%\begin{center}
%\epsfxsize=7.5 cm \epsfysize=7.5 cm \epsfbox{lattice-phase.eps} \
%\end{center}
%\caption{Sketch plot of phase diagram in $\mu_I-T$ plane(taken from \cite{Brandt:2017oyy})(Panel.(a)) and . }
%\label{lattice-phase}
%\end{figure}

Besides the above traditional methods, it is also quite interesting to study the effects of isospin chemical potential in a new framework named 'holography', which is developed based on the discovery of anti-de Sitter/conformal
field theory (AdS/CFT) correspondence\cite{Maldacena:1997re,Gubser:1998bc,Witten:1998qj}. The holographic framework has provided a new approach to deal with the non-perturbative physics of QCD, e.g. hadron spectra and hot/dense QCD matter(for details, please refer to \cite{Aharony:1999ti,Erdmenger:2007cm,deTeramond:2012rt,Adams:2012th,CasalderreySolana:2011us}). On QCD matter at finite isospin chemical potential, many efforts have been made in \cite{Albrecht:2010eg,Lee:2013oya,Nishihara:2014nva,Nishihara:2014nsa,Mamedov:2015sha}, as well. In the framework of bottom-up holographic approach, the authors of \cite{Nishihara:2014nva} studied the isospin chemical dependence of pion condensation in the framework of  hard-wall model\cite{Erlich:2005qh}. It is found that pion condensation becomes finite exactly at $\mu_I=m_\pi$ and grows with the increasing of $\mu_I$, which is consistent with the $\chi$PT investigation. Moreover, it is found chiral symmetry breaking is enhanced by the phase transition between pion condensed phase and normal phase. Compared to hard-wall model, the soft-wall model\cite{Karch:2006pv} contains the information of linear confinement as well as chiral symmetry breaking. Its extended models \cite{Gherghetta-Kapusta-Kelley,Gherghetta-Kapusta-Kelley-2,YLWu,YLWu-1,Cui:2013xva,Li:2012ay,Li:2013oda,Vega:2016gip,Capossoli:2015ywa,Capossoli:2016kcr,Capossoli:2016ydo,Zollner:2017nnh,He:2013qq} could predict meson spectra in good agreement with experimental data. It is also easy to be extended to finite temperature and give good description of chiral phase transition\cite{Colangelo:2011sr,Dudal:2015wfn,Chelabi:2015cwn,Chelabi:2015gpc,Fang:2015ytf,Li:2016gfn,Li:2016smq,Bartz:2016ufc,Fang:2016nfj,Bartz:2017jku,Fang:2018vkp,Chen:2018msc,Rodrigues:2018pep} (see also other bottom-up holographic models\cite{Iatrakis:2010jb,Jarvinen:2011qe,Alho:2012mh,Alho:2013hsa,Gursoy:2016ofp,Evans:2016jzo}). Thus, we will follow the idea of \cite{Nishihara:2014nva} and introduce temperature $T$ and isospin chemical potential $\mu_I$ in soft-wall model. Then, we will study the $T,\mu_I$  dependence of pion condensation, as well as the properties of phase transition between pion condensation phase and normal phase. For simplicity, we will focus on $N_f=2$ in this work, under which we need not to deal with more complicated phase structures related to kaon condensation phase and the other kinds of bosonic condensation.

The paper is organized as follows. In Sec.\ref{soft-intro}, we will give a brief introduction to soft-wall AdS/QCD model. We will introduce isospin density to the model and derive the equation of motion in this situation. Then, in Sec.\ref{numerical-results}, we will discuss the numerical results. We will show the isospin effects at different temperatures. Then, we will give the phase boundary of pion condensation phase. Finally, we will give a brief summary in Sec.\ref{sum}.

\section{Soft wall model with finite isospin chemical potential}
\label{soft-intro}

In bottom-up holographic framework, the soft-wall model \cite{Karch:2006pv} provides a good start point to describe both chiral symmetry breaking and linear confinement in the vacuum. In the extended soft-wall models\cite{Gherghetta-Kapusta-Kelley,Gherghetta-Kapusta-Kelley-2,YLWu,YLWu-1,Cui:2013xva,Li:2012ay,Li:2013oda,Vega:2016gip,Capossoli:2015ywa,Capossoli:2016kcr,Capossoli:2016ydo,Zollner:2017nnh,He:2013qq}, meson spectra consistent with experimental data have been obtained. At finite temperature, it has been shown in \cite{Colangelo:2011sr,Dudal:2015wfn,Chelabi:2015cwn,Chelabi:2015gpc,Fang:2015ytf,Li:2016gfn,Li:2016smq,Bartz:2016ufc,Fang:2016nfj,Bartz:2017jku,Fang:2018vkp,Chen:2018msc,Rodrigues:2018pep}  that the qualitative properties of phase transition between chiral asymmetric phase and chiral symmetric phase agree very well with the 4D understanding from lattice simulations and model calculations. Since the soft-wall model promotes the 4D global chiral symmetry $SU(N_f)_L\times SU(N_f)_R$ to 5D gauge symmetry. The 5D gauge field could be dual to the corresponding conserved current in 4D theory. Thus, it is also quite interesting to investigate the phase transition with finite isospin chemical potential $\mu_I$ by introducing $\mu_I$ through the corresponding current.   Considering only $u,d$ quarks($N_f=2$), the action of soft-wall model with $SU(2)_L\times SU(2)_R$ gauge symmetry takes the form\cite{Karch:2006pv}
\begin{eqnarray}
 S=-\int d^5x e^{-\Phi(z)} \sqrt{g}Tr\Big(D_M X^{+} D^M X+V(|X|)
 +\frac{1}{4g_5^2}(F_L^2+F_R^2)\Big), \label{action-sw}
\end{eqnarray}
with $g$ the determinant of the metric $g_{MN}$, $\Phi(z)$ the dilaton field depending only on the  fifth dimension $z$, $X$ a complex $2\times 2$ matrix-valued scalar field, $V(|X|)$ the scalar potential, $g_5=2\pi$\cite{Erlich:2005qh} the 5D gauge coupling. Since we will not consider the baryon number density, the gauge field related to $U(1)$ gauge symmetry(4D global $U_B(1)$ symmetry) is neglected. Also, since the $\omega$ meson will not be considered below, we do not have to add the Chern-Simons term in the current model(see discussion in \cite{Nishihara:2014nva}). In the above action, $F_L$ and $F_R$ are the field strength tensor of left and right gauge field $L_M, R_M$, defined as
\begin{eqnarray}
F_{L,MN}&=&\partial_{M}{L_{N}}-\partial_{N}{L_{M}}-i[L_{M},L_{N}],\nonumber\\
F_{R,MN}&=&\partial_{M}{R^{N}}-\partial_{N}{R_{M}}-i[R_{M},R_{N}].
\end{eqnarray}
Accordingly, the covariant derivative $D_M$  will be defined as
\begin{equation}
D_M X=\partial_M X-i L_M X+iX R_M.
\end{equation}
One can redefine the vector ($V$) and axial-vector ($A$) gauge fields as
\begin{eqnarray}
V_M=\frac{L_M+R_M}{2}, A_M=\frac{L_M-R_M}{2}.
\end{eqnarray}
Then one has $F_{L}^{2}+F_{R}^{2} = 2 \left(F_{V}^{2}+  F_{A}^{2}\right)$, with
\begin{eqnarray}
F_{V,MN}&=&\partial_{M}{V_{N}}-\partial_{N}{V_{M}}-\frac{i}{\sqrt{2}}[V_{M},V_{N}],\\
F_{A,MN}&=&\partial_{M}{A_{N}}-\partial_{N}{A_{M}}-\frac{i}{\sqrt{2}}[A_{M},A_{N}].
\end{eqnarray}
The covariant derivative becomes
\begin{equation}
D_M X=\partial_M X-i[V_{M},X]-i\{A_{M},X\}.
\end{equation}
Here $V_M, A_M$ are matrix-valued gauge filed as $V_M=V_M^a t^a, A_M=A_M^a t^a, a=1,2,3$, with $t^a$ the generator of $SU(2)$ group satisfying Tr$[t^{a}t^{b}]=\delta^{ab}/2$.
Generally, the eight real degrees of freedom of the $2\times2$ complex matrix-valued scalar field could be parameterized as
\begin{eqnarray}
X=(\chi t^0 +S^a t^a)e^{i2\pi^b t^b+i\eta}
\end{eqnarray}
with $a,b=1,2,3$, $t^0=\frac{I_2}{2}$ and $I_2$ the $2\times2$ identity matrix. Using the gauge symmetry, one can take the fifth component of gauge fields to be zero, i.e. $V_z=A_z=0$.

If one focuses on finite temperature case, then both the vector and axial vector field vanish. In this case, one can check $S^a, \pi^b,\eta$ vanish if there are no extra source of the corresponding operator. The phase structure in this situation has been studied in \cite{Colangelo:2011sr,Dudal:2015wfn,Chelabi:2015cwn,Chelabi:2015gpc,Fang:2015ytf,Li:2016gfn,Li:2016smq,Bartz:2016ufc,Fang:2016nfj,Bartz:2017jku,Fang:2018vkp,Chen:2018msc,Rodrigues:2018pep}  . In this work, since we would like to discuss the effects of isospin chemical potential on phase transition, we have to take non-vanishing $V^M, A^M$. According to the discussion of \cite{Nishihara:2014nsa,Nishihara:2014nva}, with finite isospin chemical potential,
the non-vanishing components of scalar field and gauge filed could be set as
$\chi, \Pi\equiv\pi^1, V_0^3, A_0^1, A_0^2$. Therefore, the non-vanishing field contents left  are as following
\begin{eqnarray}
X=\chi (t^0 \cos \Pi + i t^1 \sin\Pi), A_0^1, A_0^2, V_0^3.
\end{eqnarray}
If we only consider the homogeneous phase, we can assume that all the above fields depend only on the fifth coordinate $z$. In \cite{Nishihara:2014nsa,Nishihara:2014nva}, the authors consider isospin chemical potential effect and take the metric ansatz as the pure $AdS_5$ metric(in this work, we will always take the AdS radius as $1$)
\begin{eqnarray}
ds^2=\frac{1}{z^2}(-dt^2+dz^2+dx^i dx_i).
\end{eqnarray}
The temperature effect has not been taken into account. Here, we would like to study the temperature effect also. Thus, we take the following metric ansatz
\begin{eqnarray}
ds^2=e^{2A(z)}(-f(z) dt^2+\frac{1}{f(z)}dz^2+dx_i dx^i).
\end{eqnarray}
The temperature could be introduced if there is a horizon $z=z_h$ where $f(z)=0$. The temperature is related to $z_h$ by the formula
\begin{eqnarray}
T=|\frac{f^{'}(z_h)}{4\pi}|.
\end{eqnarray}
In general, $A(z),f(z)$ should be solved from certain kind of gravity system coupled with the soft-wall model action. But the full back-reaction solution is difficult to obtained,  and the simple approximation 'potential reconstruction method' used in \cite{Li:2011hp,Kajantie:2011nx,Yang:2014bqa,Li:2017tdz,Chen:2017cyc,Li:2014hja,Li:2014dsa,Cai:2012xh,Li:2017ple} can not be used here, because it can not take into account the temperature dependent of condensations. Thus, for simplicity, in the sense of probe limit, we take the following Anti-de Sitter-Reissner-Nordstrom (AdS-RN) metric solution with finite isospin number
\begin{eqnarray}
A(z)&&=-\ln(z),\\
f(z)&&=1-(1+\gamma \mu_I^2 z_h^2)\frac{z^4}{z_h^4}+\gamma \mu_I^2 \frac{z^6}{z_h^4},\\
v&&\equiv V_0^3(z)=\mu_I(1-\frac{z^2}{z_h^2}).
\end{eqnarray}
Here, $\mu_I$ is the isospin chemical potential and $\gamma$ is related to the coupling of $V_0^3$ with gravity, which will be taken as a free parameter. For simplicity, we take $\gamma=1$ in later calculation. Different from \cite{Nishihara:2014nsa,Nishihara:2014nva}, we will take $V_0^3$ as a background field other than a dynamical field in the following discussion. Under this background, temperature becomes a simple function of the horizon $z_h$
\begin{eqnarray}
T=\frac{2-\gamma \mu_I^2 z_h^2}{2\pi z_h},
\end{eqnarray}
in which we have taken the outer horizon $z_h<\frac{\sqrt{2}}{\sqrt{\gamma}\mu_I}$. We note that at finite isospin chemical potential it is $z_h=\frac{\sqrt{2}}{\sqrt{\gamma}\mu_I}$ other than $f\equiv1$, standing for $T=0$ background\footnote{Here, we would like to emphasize that the zero temperature limit of the AdS-RN metric does not lead $f(z)\equiv1$ at finite $\mu_I$. Thus, the metric ansatz in \cite{Nishihara:2014nva} is quite different from that of this work. } .

According to the above discussion, effectively, the 5D Lagrangian becomes
\begin{eqnarray}
\mathcal{L}_{eff}=&&-\frac{e^{3A-\Phi}f}{2}[\chi^{'2}+\chi^2 \Pi^{'2}]+\frac{e^{A-\Phi}}{2g_5^2}[a_1^{'2}+a_2^{'2}]-e^{5A-\Phi}V(\chi)\nonumber\\
         && +\frac{e^{3A-\Phi}\chi^2}{2f}[v^2\sin^2(\Pi)-a_2 v \sin(2\Pi)+a_1^2+a_2^2\cos^2(\Pi)].
\end{eqnarray}
Here, we have replaced $V_0^3,A_0^1, A_0^2$ with more compact form $v\equiv V_0^3, a_1=A_0^1, a_2=A_0^2$, and redefined $V(\chi)\equiv \text{Tr}[V(|X|)]$. As \cite{Chelabi:2015cwn,Chelabi:2015gpc}, we will take
\begin{eqnarray}
V(\chi)=-\frac{3}{2}\chi^2+v_4 \chi^4.\label{Vx}
\end{eqnarray}
One can easily get the equation of motion for $a_1$ as
\begin{eqnarray}
\partial_z(\frac{e^{A-\Phi}}{g_5^2}\partial_z a_1)-\frac{e^{3A-\Phi}\chi^2}{f}a_1=0.
\end{eqnarray}
Since we do not have explicit source for the axial-vector current, thus we require $a_1=0$ at the boundary. Furthermore, the background metric considered is a black hole solution. Thus, $a_1(z_h)=0$ is also necessary. One can easy check that $a_1=0$ satisfies both the equation of motion and the boundary conditions. Thus, we will take $a_1=0$ as an ansatz, which is different from the study in \cite{Nishihara:2014nsa,Nishihara:2014nva}. Then, it is also quite direct to derive the equation of motion for the corresponding fields $\chi,\Pi,a_2$ as
\begin{eqnarray}
&&\partial_z(e^{3A-\Phi}f\partial_z \chi)-e^{5A-\Phi}\partial_\chi V(\chi)\nonumber\\
&&-\frac{e^{3A-\Phi}\chi}{f}(f^2\Pi^{'2}-v^2\sin^2(\Pi)+a_2 v \sin(2\Pi)-a_2^2\cos^2(\Pi))=0\label{eq-chi}
\end{eqnarray}
\begin{eqnarray}
\partial_z(e^{3A-\Phi}f\chi^2\partial_z \Pi)+\frac{e^{3A-\Phi}\chi^2}{2f}[(v^2-a_2^2)\sin(2\Pi)-2a_2 v \cos(2\Pi)]=&&0\label{eq-pi}\\
\partial_z(\frac{e^{A-\Phi}}{g_5^2}\partial_z a_2)-\frac{e^{3A-\Phi}\chi^2}{2f}[2a_2 \cos^2(\Pi)-v\sin(2\Pi)]=&&0\label{eq-a2}
\end{eqnarray}

\begin{figure}[h]
\begin{center}
\epsfxsize=4.0 cm \epsfysize=4.0 cm \epsfbox{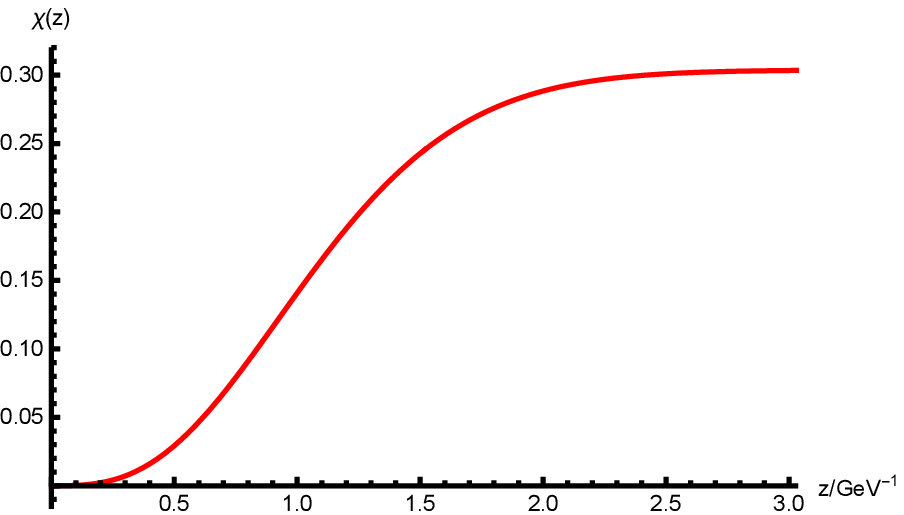}
\hspace*{0.1cm} \epsfxsize=4.0 cm \epsfysize=4.0 cm \epsfbox{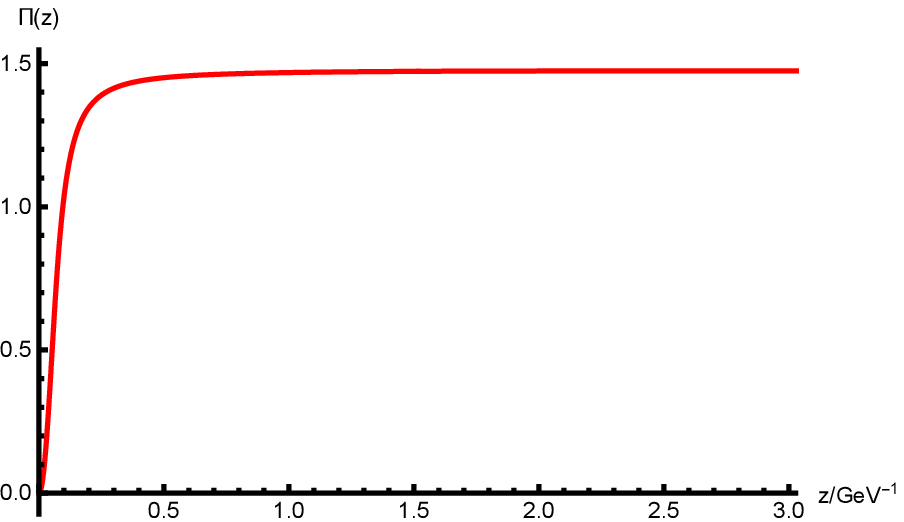}
\hspace*{0.1cm} \epsfxsize=4.0 cm \epsfysize=4.0 cm\epsfbox{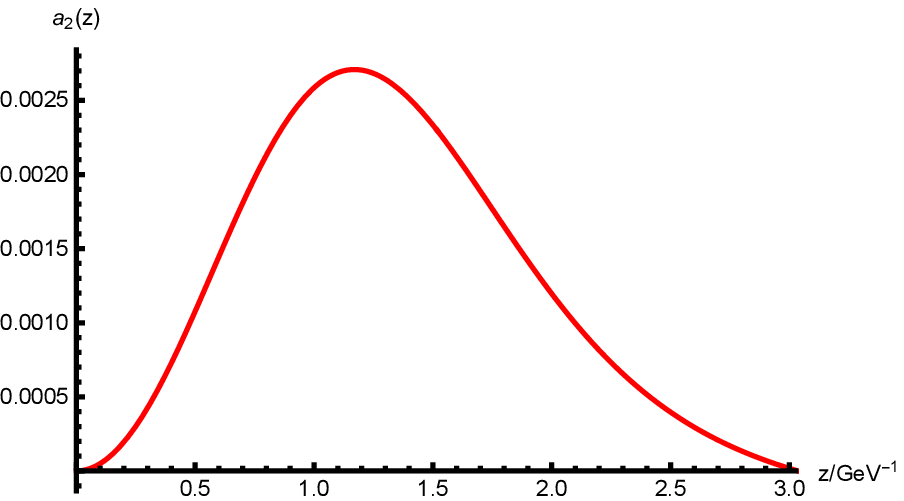}
\vskip -0.05cm \hskip 0.15 cm
\textbf{( a ) } \hskip 4.0 cm \textbf{( b )} \hskip 4.0 cm \textbf{( c )} \\
\end{center}
\caption{Solution with $T=100\rm{MeV}, \mu=100\rm{MeV}, m=5\rm{MeV},$ $ v_4=8$ and $\Phi(z)=-z^2$. The integral constants are solved as $\sigma=0.007\rm{GeV}^3,\pi_1=0.077\rm{GeV}^3,a_{2u}=0.005\rm{GeV}^2,\chi_0=0.303,\Pi_0=1.475,d_1=-0.0005\rm{GeV}$. }
\label{sol-eg}
\end{figure}

This group of equations are too complicated to impose analytical analysis. To get the numerical solution, firstly, we have to make the boundary conditions clear.  At the UV boundary $z=0$, the leading expansion of $\chi, \Pi, a_2$ could be solved as
\begin{eqnarray}
\chi(z)&&=m \zeta z+ m \zeta \Phi^{'}(0) z^2+\frac{1}{2}m\zeta(\Phi^{''}(0)-2\Phi^{'2}(0)+4 v_4 m^2\zeta^2)z^3\ln(z)+\frac{\sigma}{\zeta}z^3+o(z^3),\nonumber\\
\Pi(z)&&=\frac{\pi_1}{m\zeta^2} z^2+2\frac{\pi_1}{m\zeta^2}\Phi^{'}(0)z^3+o(z^3),\nonumber\\
a_2(z)&&=a_{2u}z^2+\frac{2}{3}a_{2,u}\Phi^{'}(0)z^3,
\end{eqnarray}
where we have imposed the AdS/CFT dictionary and identified the coefficient of the leading terms with quark mass $m$, chiral condensate $\sigma$, pion condensate $\pi_1$. In this expression, we have neglected the constant term in $\Pi(z)$ and $a_2(z)$, since they are related to the sources of axial current, which is not be considered in this work. The constant $\zeta=\frac{\sqrt{N_c}}{2\pi}$ is introduced to match the two point function $\bar{q}q(p)\bar{q}{q}(0)$ from holographic calculation and 4D calculation(For details, please refer to \cite{Cherman:2008eh}). Similarly, one can calculate the near horizon expansion as
\begin{eqnarray}
\chi(z)&&=\chi_0-\frac{3\chi_0-4v_4\chi_0^3}{2z_h(-2+\gamma\mu^2z_h^2)}(z-z_h)+o(z-z_h)\nonumber\\
\Pi(z)&&=\Pi_0+\frac{-4d_1 \mu z_h\cos(2\Pi_0)+(d_1^2 z_h^2-4\mu^2)\sin(2\Pi_0)}{32(2z_h(-2+\gamma\mu^2z_h^2)}(z-z_h)+o(z-z_h)\nonumber\\
a_2(z)&&=d_1(z-z_h)+o(z-z_h)
\end{eqnarray}
These UV and IR expansions guarantee the AdS/CFT dictionary and the regularity of physical solution. The UV coefficient $m$ could be fixed by the quark mass considered, which in some sense could be tuned as a parameter\footnote{Due of color confinement, the exact values of current quark masses are not measured directly.}. The other integral constants $\sigma,\pi_1,a_{2u}, \chi_0, \Pi_0, d_1$ could be solved self-consistently using 'shooting-method', requiring the solutions from UV boundary and IR boundary connecting smoothly in the intermediate region. We take $T=100\rm{MeV}, \mu_I=100\rm{MeV}, m=5\rm{MeV},$ $v_4=8$ and $\Phi(z)=-z^2$  as an example, and show the solution in Fig.\ref{sol-eg}. From Fig.\ref{sol-eg}, we could see that  all the three solutions $\chi,\Pi,a_2$ equals zero at UV boundary. At the horizon $z_h=3.04\rm{GeV}^{-1}$, $\chi=0.303,\Pi=1.475$, while $a_2=0$. Also, at this condition, both chiral condensation and pion condensation are finite.

\section{Chiral condensation and pion condensation at finite isospin chemical potential}
\label{numerical-results}

In last section, we have given a short introduction on how to introduce isospin chemical potential in soft-wall model. In this section, we will  continue the discussion and calculate the temperature and isospin dependence of both chiral condensation and pion condensation. In \cite{Chelabi:2015gpc,Li:2016gfn,Li:2016smq,Chen:2018msc}, we have shown that the qualitative properties of chiral phase transition could be well described when the dilaton profile takes the following form
\begin{eqnarray}\label{intdilaton}
\Phi(z)=-\mu_1^2z^2+(\mu_1^2+\mu_0^2)z^2\tanh(\mu_2^2z^2).
\end{eqnarray}
At UV, $\Phi(z)$ tends to $-\mu_1^2 z^2$, which is responsible for chiral symmetry spontaneous breaking as shown in \cite{Chelabi:2015gpc}. In addition $\Phi(z)$ tends to $\mu_0^2z^2$ when $z\rightarrow \infty$, which is responsible for the linear spectra. Considering the vacuum value of chiral condensate, the transition temperature of chiral phase transition and the Regge slope of meson spectra, we will follow our previous study and take $\mu_0=0.430 \rm{GeV},\mu_1=0.830\rm{GeV}, \mu_2=0.176\rm{GeV}$ in the following calculation. Further, for the parameter $v_4$ in the scalar potential Eq.(\ref{Vx}), we will also follow our previous fitting and take $v_4=8$.  For a summary, we list all the parameter values in Table.\ref{parameters}. Instead of trying to get quantitative result, in this work, we will focus on study the qualitative behavior of phase transition between pion condensation phase and normal phase, as well as chiral phase transition. Inserting the dilaton profile Eq.(\ref{intdilaton}) and the scalar potential Eq.(\ref{Vx}) into the equations of motion Eqs.(\ref{eq-chi},\ref{eq-pi},\ref{eq-a2}), we can solve out $\sigma, \pi_1$, as well as the other integral constants. Since the equations of motion are quite complicated, analytical analysis is impossible. So we use numerical method to solve it. From the numerical results, solution structure at finite isospin chemical potential depends on temperature. Solutions with finite pion condensation only exist at temperature below $T_3=0.127\rm{GeV}$. However, comparing the free energy, we find that only below $T_2=0.113\rm{GeV}$, there could be thermodynamical stable solutions with finite pion condensation. In the following sections, we will describe the details of numerical results obtained from the soft-wall model given above.
\begin{table}
\begin{center}
\begin{tabular}{cccccccc}
\hline\hline
parameter & $\gamma$   &   $v_4$   &   $\mu_0$         &  $\mu_1$          &    $\mu_2$            \\
\hline
value     &  $1$      &   $8$   & $0.430\rm{GeV} $  &  $0.830\rm{GeV}$   &   $0.176\rm{GeV}$  \\
\hline
\end{tabular}
\caption{The values of free parameters used in numerical calculation.}
\label{parameters}
\end{center}
\end{table}

\subsection{Low temperature results}
\begin{figure}[h]
\begin{center}
\epsfxsize=4.5 cm \epsfysize=4.5 cm \epsfbox{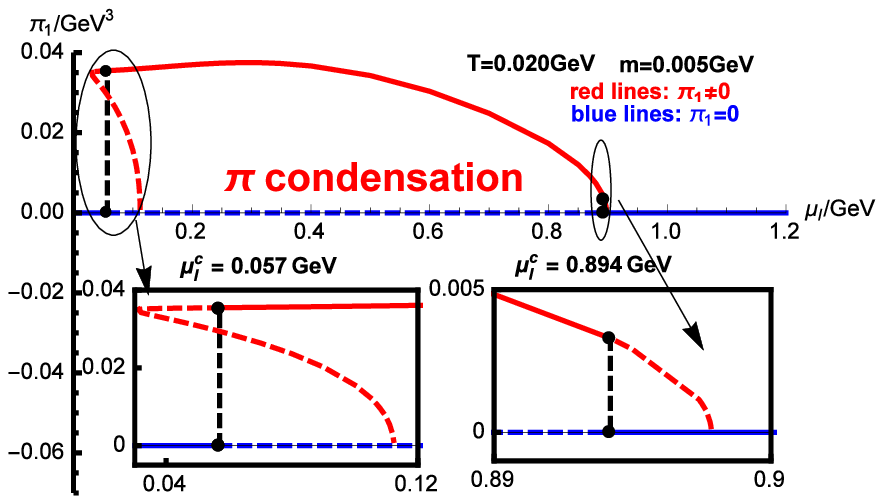}
\hspace*{0.1cm} \epsfxsize=4.5 cm \epsfysize=4.5 cm \epsfbox{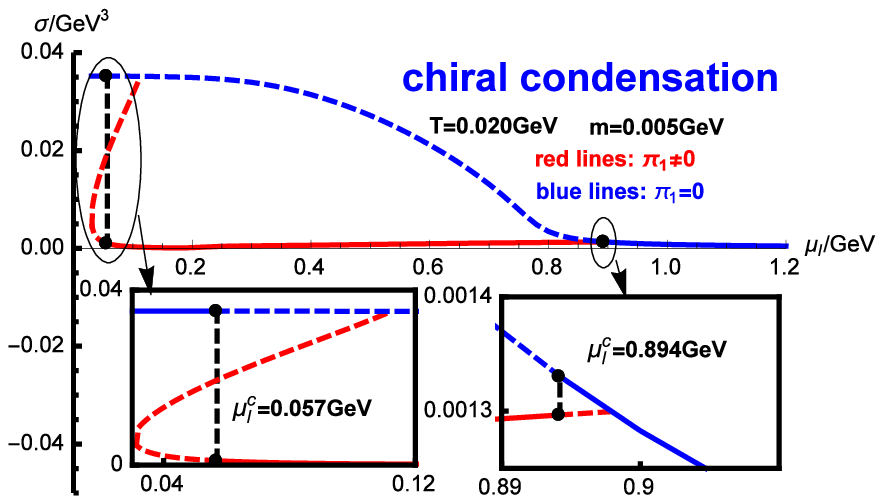}
\hspace*{0.1cm} \epsfxsize=4.5 cm \epsfysize=4.5 cm \epsfbox{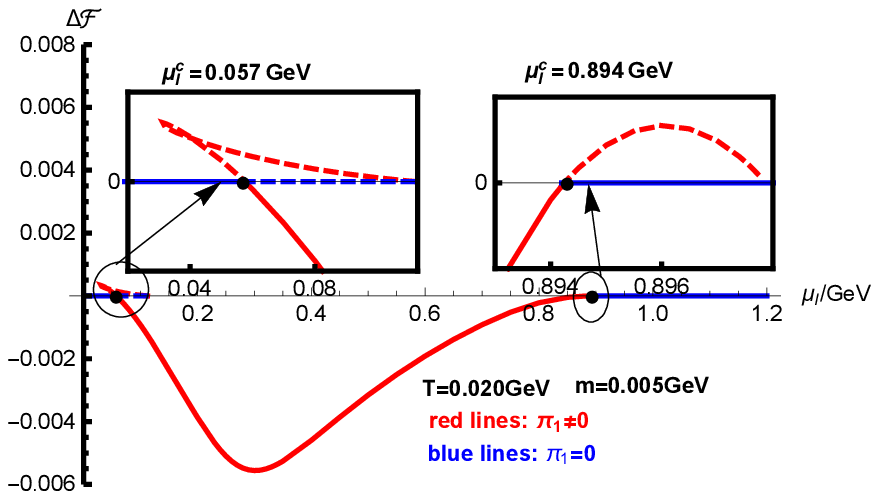}
\vskip -0.05cm \hskip 0.15 cm
\textbf{( a ) } \hskip 4.5 cm \textbf{( b )} \hskip 4.5 cm \textbf{( c )}\\
\end{center}
\caption{Pion condensation(Panel.(a)), chiral condensation(Panel.(b)) and free energy difference $\Delta\mathcal{F}$(Panel.(c)) as functions of isospin chemical potential $\mu_I$ at $T=0.020\rm{GeV}$. Qualitatively, the solution structures for $T<0.080\rm{GeV}$ are similar to the one showed here. The red and blue lines in Panels.(a),(b) and (c) represent results with and without pion condensation respectively. The red and blue solid lines represent the thermodynamical stable solutions, while the red and blue dashed lines represent thermodynamical unstable solutions. The black dashed lines are auxiliary line to show the jump between different solutions, which shows a first order phase transition. The embedded figures are the enlarge view of the corresponding areas.    }
\label{T020}
\end{figure}

Firstly, we investigate the isospin chemical potential effect at low temperature.  We find that below $T_1=0.080\rm{GeV}$, the characteristic behavior of the solutions is like the results for $T=0.020\rm{GeV}$, which is given in Figs.\ref{T020}.

It is easy to see from Eqs.(\ref{eq-chi},\ref{eq-pi},\ref{eq-a2}) that if one sets $\pi_1=0,a_{2,u}=0$, then the equations reduced to the one without pion condensation. In this situation, depending on the dilaton profile, chiral condensate could be finite. From our numerical analysis, we found that there are always solutions with finite chiral condensate and vanishing pion condensate. The results are given in the blue lines(both dashed and solid ones) in Figs.\ref{T020}.  However, if one hopes to get solutions with finite pion condensation, then it is necessary to solve the coupled equations Eqs.(\ref{eq-chi},\ref{eq-pi},\ref{eq-a2}). We find that for $T=0.020\rm{GeV}$, only at the range of $0.313\rm{GeV}<\mu_I<0.897\rm{GeV}$, there are solutions with finite pion condensation. At $\mu_I=0.897\rm{GeV}$, pion condensation disappears, and the solutions with and without pion condensation merge together. This could be seen from Fig.\ref{T020} in the enlarge view of the corresponding area. From the enlarge view in Fig.\ref{T020}(a), we could see that $\pi_1$ decreases monotonically to zero. From a first glance, there should be a second order phase transition in this region, because the order parameter $\pi_1$ decreases continuously from finite to zero.

To check the order of phase transition, we calculate the free energy according to the equivalence of the partition function conjecture
\begin{eqnarray}
Z_{QCD}=Z_{gravity}\backsimeq e^{-S_E}=e^{-\beta F},
\end{eqnarray}
which gives the expression of the free energy density
\begin{eqnarray}
\mathcal{F}\equiv \frac{F}{V_3}=-\int_0^{z_h} dz \mathcal{L}_{eff}.\label{freeenergy}
\end{eqnarray}
Here $F$ is the free energy, $S_E$ is the Euclidean on-shell action and $V_3$ is the three dimension volume.  In principle, one can insert the solutions obtained above and get the free energy. However, when $m\neq 0$, the above expression is divergent around $z=0$. Thus, holographic re-normalization process are required for finite quark masses to get finite free energy density. Noting that the main goals of calculating free energy here is to determine the thermodynamical favored solution at the same $\mu_I$ and $T$, the real quantity needed is the free energy differences
\begin{eqnarray}
\Delta\mathcal{F}=\mathcal{F}_{\pi\neq0}-\mathcal{F}_{\pi=0}.\label{dfree}
\end{eqnarray}
In probe limit, the background metrics are the same for different solutions with the same $\mu_I, T$. Therefore, one just inserts the solutions into Eqs.(\ref{freeenergy},\ref{dfree}). Then it is easy to see that the divergence is cancelled for solutions with the same $\mu_I, T, m$. Then one does not need to do holographic re-normalization procedure.

Following this logic, we calculate the free energy differences of solutions with and without pion condensation. The results are shown in Fig.\ref{T020}(c). In Fig.\ref{T020}(c),  from the enlarge view of the region around $\mu_I=0.894\rm{GeV}$, before pion condensation decreases to zero at $\mu_I^c=0.897\rm{GeV}$, $\Delta \mathcal{F}$ reaches zero at $\mu_I^c=0.894\rm{GeV}$, revealing the thermodynamically instability of pion condensation above $\mu_I^c$(labeled with dashed lines). Thus, though it looks like a second order phase transition in terms of the behavior of pion condensation, the real transition order is first order.

\begin{figure}[h]
\begin{center}
\epsfxsize=6.5 cm \epsfysize=6.5 cm \epsfbox{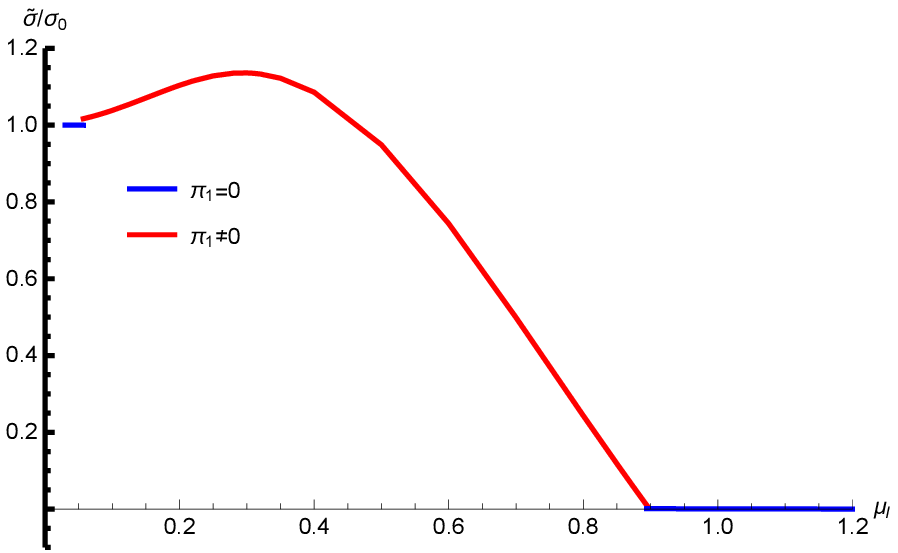}
\hspace*{0.1cm} \epsfxsize=6.5 cm \epsfysize=6.5 cm \epsfbox{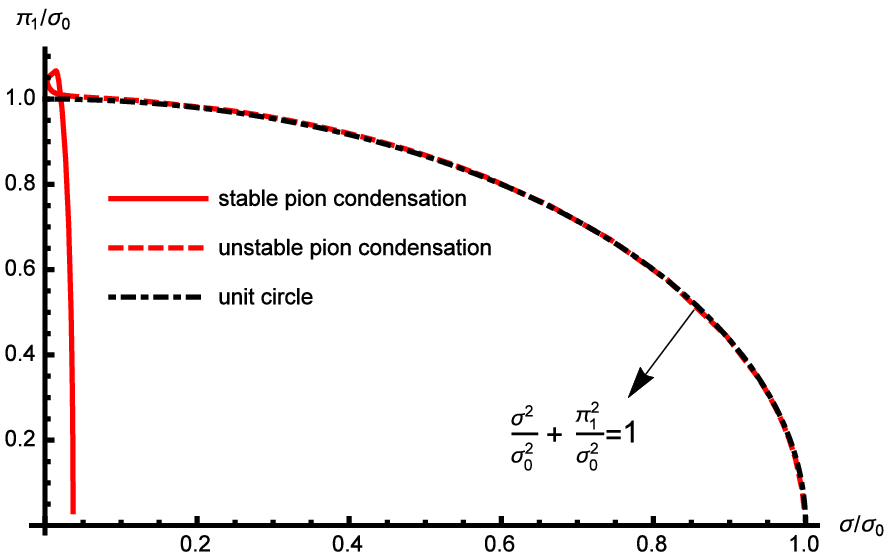}
\vskip -0.05cm \hskip 0.15 cm
\textbf{( a ) } \hskip 6.5 cm \textbf{( b )} \\
\end{center}
\caption{The full condensation(Panel.(a)) $\tilde{\sigma}\equiv\sqrt{\sigma^2+\pi_1^2}$ and chiral circle (Panel.(b))  as functions of isospin chemical potential $\mu_I$ at $T=0.020\rm{GeV}$.  The red lines represent solution with pion condensation and the blue lines represent solutions without.  }
\label{T020-circle}
\end{figure}

Then, we turn to the small $\mu_I$ region. At around $0.0313\rm{GeV}<\mu_I<0.113\rm{GeV}$, there are three solutions for each value of $\mu_I$. Two of them have finite pion condensation, while the other one without pion condensation(for details, please refer to the enlarge view of the corresponding area). The triple solution structure is a characterize signal of first order phase transition. To confirm this, we also calculate the free energy differences and show the results in Fig.\ref{T020}(c). From Fig.\ref{T020}(c), we could see a swallowtail structure in the enlarge view. This confirms a first order phase transition at $\mu_I^c=0.057\rm{GeV}$.

From the above discussion, the thermodynamical stable path are the solid lines(both red and blue lines) in Fig.\ref{T020}. At $\mu_I<0.057\rm{GeV}$, there are no pion condensation and chiral condensation equals almost to its vacuum value $\sigma_0=0.0352\rm{GeV}^3=(327\rm{MeV})^3$. At $\mu_I=0.057\rm{GeV}$, a first order phase transition happens and pion condensation jumps to a finite value $\pi_1=0.0355\rm{GeV}^3\approx 1.01\sigma_0$. The system transits from normal phase to pion condensed phase. Meanwhile, chiral condensate decreases from its vacuum value $\sigma_0$ to a very small value $\sigma=0.001\rm{GeV}^3\approx0.03\sigma_0$. It seems that chiral symmetry is restored together with the transition to pion condensed phase. However, since isospin chemical potential will rotate the condensates, when considering chiral symmetry, one has to considered the full condensate $\tilde{\sigma}\equiv\sqrt{\sigma^2+\pi_1^2}$. To check the restoration of chiral symmetry, we plot $\tilde{\sigma}$ as a function of $\mu_I$ in Fig.\ref{T020-circle}(a). From the plot, we could see that instead of decreasing to a small value, $\tilde{\sigma}$ jumps to a larger value, showing the enhancement of chiral symmetry after the phase transition at $\mu_I=0.057\rm{GeV}$. Despite of the transition order, the enhancement of chiral symmetry after the phase transition is similar to the study in hard-wall model\cite{Nishihara:2014nsa,Nishihara:2014nva}. When $\mu_I$ continues to increase, pion condensation increases continuously to a maximal value  $\pi_1=0.0375\rm{GeV}\approx 1.07\sigma_0$ at $\mu_I=0.294\rm{GeV}$, and then starts to decrease until the phase transition at $\mu_I=0.894\rm{GeV}$. During this procedure, $\tilde{\sigma}$ also increase first and then decrease to almost zero, showing enhancement of chiral symmetry breaking at small value of $\mu_I$ and suppression of chiral symmetry breaking at very large $\mu_I$. After the phase transition, both chiral symmetry is restored and pion condensation disappears. The system become symmetry restored phase and normal phase without pion condensation. At low temperature, we have seen more complicated phase structure than $\mu_I=0$ case.

Finally, we also show the chiral circle in Fig.\ref{T020-circle}(b). From the figure, we find that points of $(\sigma_0,\pi_1)$ at small $\mu_I$ locate almost at the unit circle $\frac{\sigma^2}{\sigma_0^2}+\frac{\pi_1^2}{\sigma_0^2}$. This is in agreement with the study of hard-wall calculation in \cite{Nishihara:2014nsa,Nishihara:2014nva}. Nevertheless, this part of solutions is unstable. Instead, the stable solutions in the intermediate region of $\mu_I$ deviate from the unit circle.

\subsection{High temperature results}

\begin{figure}[h]
\begin{center}
\epsfxsize=4.5 cm \epsfysize=4.5 cm \epsfbox{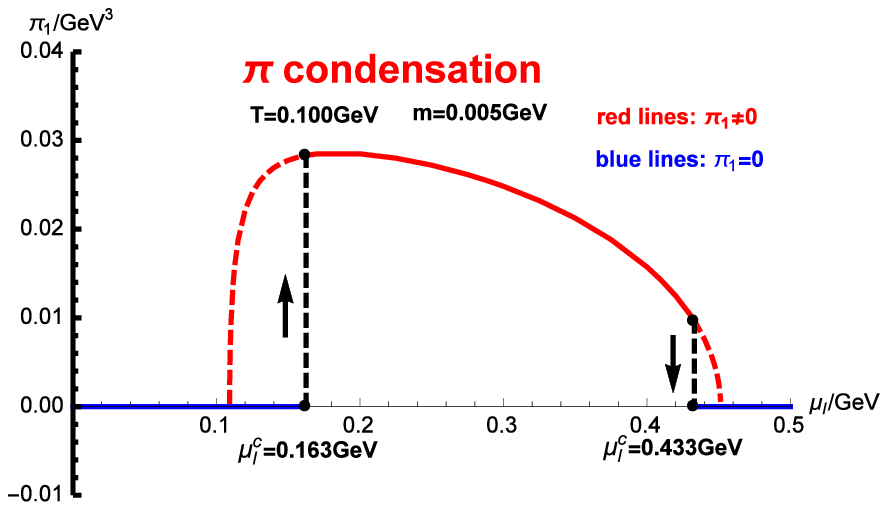}
\hspace*{0.1cm} \epsfxsize=4.5 cm \epsfysize=4.5 cm \epsfbox{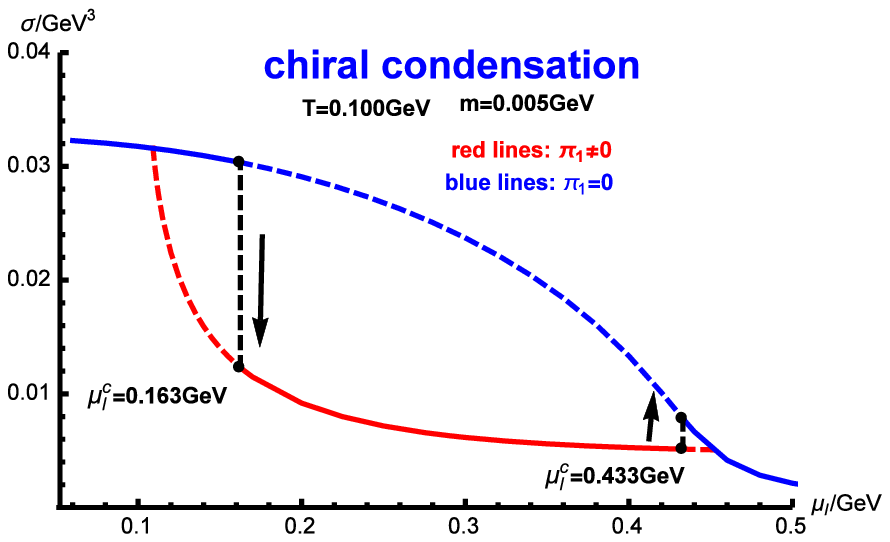}
\hspace*{0.1cm} \epsfxsize=4.5 cm \epsfysize=4.5 cm \epsfbox{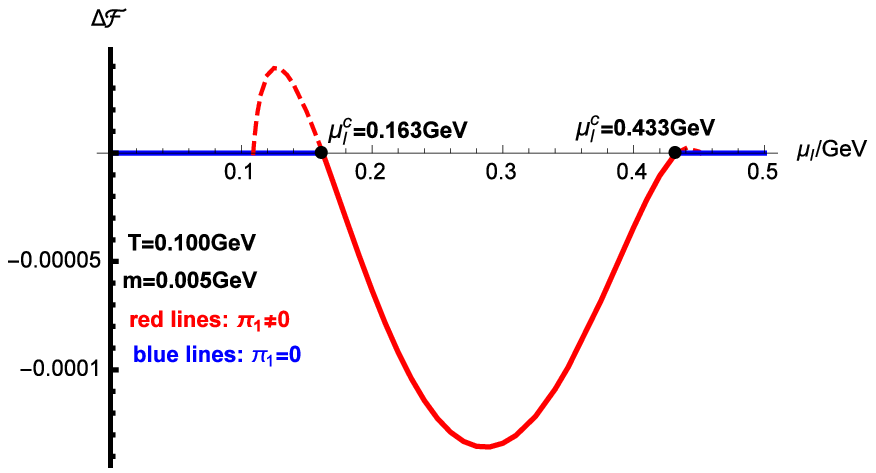}
\vskip -0.05cm \hskip 0.15 cm
\textbf{( a ) } \hskip 4.5 cm \textbf{( b )} \hskip 4.5 cm \textbf{( c )} \\
\end{center}
\caption{Condensations $\pi_1$(a),$\sigma$(b) and free energy difference $\Delta \mathcal{F}$(c) as functions of isospin chemical potential $\mu_I$ when $T=0.100\rm{GeV}$.  In both (a) and (b), red and blue lines are solution from EoMs with and without pion condensation respectively.  Solutions with pion condensation exist in the range $0.109\rm{GeV}<\mu_I<0.451\rm{GeV}$. Black solid lines are auxiliary lines to show the jump of $\pi_1,\sigma$ at the first order transition points, locating at $\mu_I^c=0.163\rm{GeV}$ and $\mu_I^c=0.433\rm{GeV}$. According to the free energy comparison, the red and blue  solid lines are thermodynamical stable solutions, while the red and blue dashed lines are thermodynamical unstable solutions. The thermodynamical favored path is along the solid lines.  }
\label{T100-cd}
\end{figure}

From our numerical analysis, we find that the small $\mu_I$ triple solution region shrinks to a point when increasing the temperature, and it ends to be a point at $T=0.080\rm{GeV}$.  When $0.080\rm{GeV}<T<0.114\rm{GeV}$, the solution structure is like the plot shown in Fig.\ref{T100-cd}, where we have taken $T=0.100\rm{GeV}$ as an example to show the qualitative behavior.

\begin{figure}[h]
\begin{center}
\epsfxsize=6.5 cm \epsfysize=6.5 cm \epsfbox{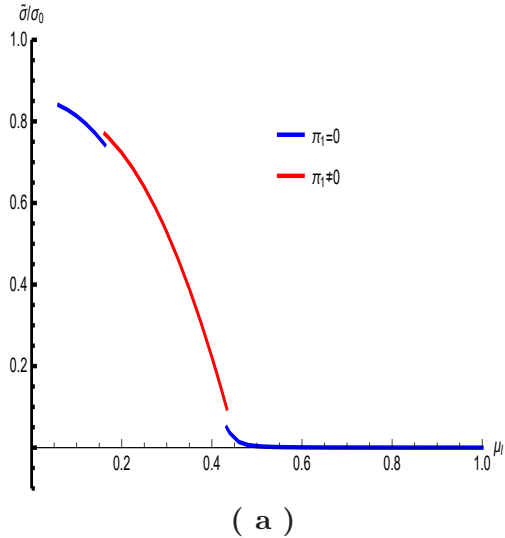}
\vskip -0.05cm \hskip 0.15 cm
\textbf{( a ) }  \\
\end{center}
\caption{The full condensation(Panel.(a)) $\tilde{\sigma}\equiv\sqrt{\sigma^2+\pi_1^2}$ as a functions of isospin chemical potential $\mu_I$ at $T=0.100\rm{GeV}$.  The red lines represent solution with pion condensation and the blue lines represent solutions without.  }
\label{T100-st}
\end{figure}

From Fig.\ref{T100-cd}(a), at $T=0.1\rm{GeV}$, there are no solutions with pion condensation when $\mu_I<0.109\rm{GeV}$. From Fig.\ref{T100-cd}(b), the solutions without pion condensation have finite chiral condensation, which decreases with the increasing of $\mu_I$. Comparing to $T=0.020\rm{GeV}$, we observe that at the same $\mu_I$ $\sigma$ of $T=0.100\rm{GeV}$ is smaller, indicating the suppression of condensate by temperature effect. Then, when $\mu_I$ increases, solutions with pion condensation appear. Different from $T=0.020\rm{GeV}$, pion condensation increase monotonically and continuously from zero. There are no triple solution regions any more. Similarly, though this behavior looks very like a second order phase transition, we have to check it with free energy comparison. Therefore, we plot the free energy difference $\Delta\mathcal{F}$ in Fig.\ref{T100-cd}(c).  From the plot, we find that the free energy of solutions with pion condensation in the range $0.109\rm{GeV}<\mu_I<0.163\rm{GeV}$ is larger than that of the solutions without pion condensation. This indicates the phase transition is a first order one. Pion condensation jumps discontinuously from zero to a finite value $\pi_1=0.0283\rm{GeV}$ at $\mu_I=0.163\rm{GeV}$. At the same time, chiral condensate decreases suddenly from $\sigma=0.0303\rm{GeV}$ to $\sigma=0.010\rm{GeV}$. Also, to check the effect on chiral symmetry restoration, we examine the full condensation $\tilde{\sigma}$ in Fig.\ref{T100-st}. From the plot, we find that $\tilde{\sigma}$ jumps to a higher value at the transition point. So the phase transition enhance chiral symmetry breaking. But different from low temperature region, $\tilde{\sigma}$ decreases monotonically after the transition until the large $\mu_I$ transition. After the phase transition, pion condensation is still not monotonic function of $\mu_I$. It increases in the range of $0.163\rm{GeV}<\mu_I<0.200\rm{GeV}$ and decreases in the range $0.200\rm{GeV}<\mu_I<0.433\rm{GeV}$. Above $\mu_I=0.433\rm{GeV}$, solutions with pion condensation have larger free energy from Fig.\ref{T100-cd}(c), and it is thermodynamically unfavored, which should be replaced by the normal phase without pion condensation. There is a first order phase transition at this point.

\subsection{The phase boundary}
\begin{figure}[h]
\begin{center}
\epsfxsize=4.5 cm \epsfysize=4.5 cm \epsfbox{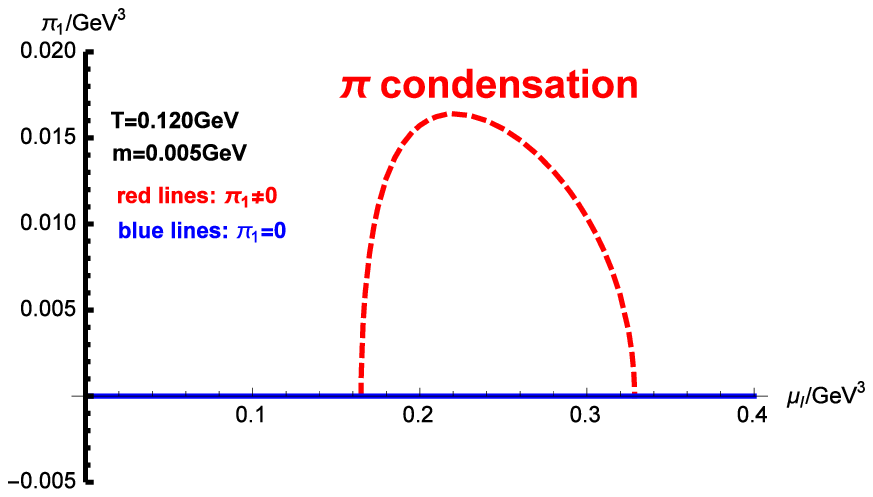}
\hspace*{0.1cm} \epsfxsize=4.5 cm \epsfysize=4.5 cm \epsfbox{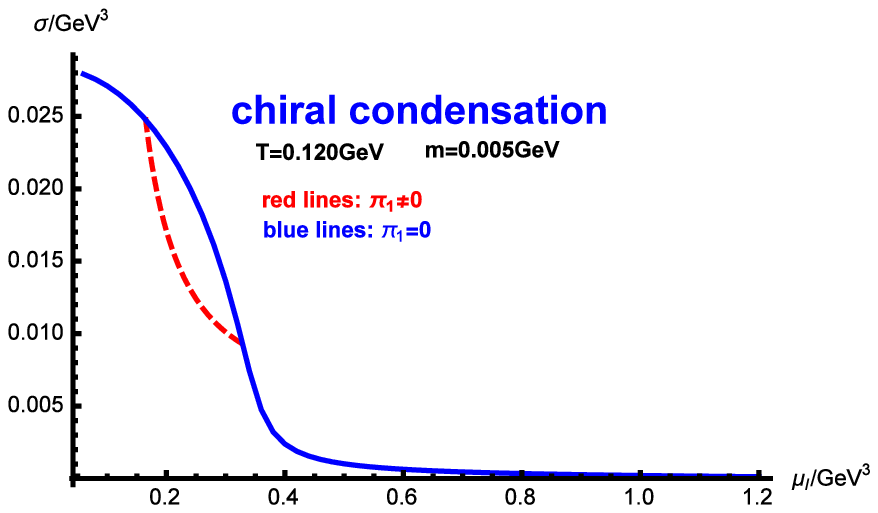}
\hspace*{0.1cm} \epsfxsize=4.5 cm \epsfysize=4.5 cm \epsfbox{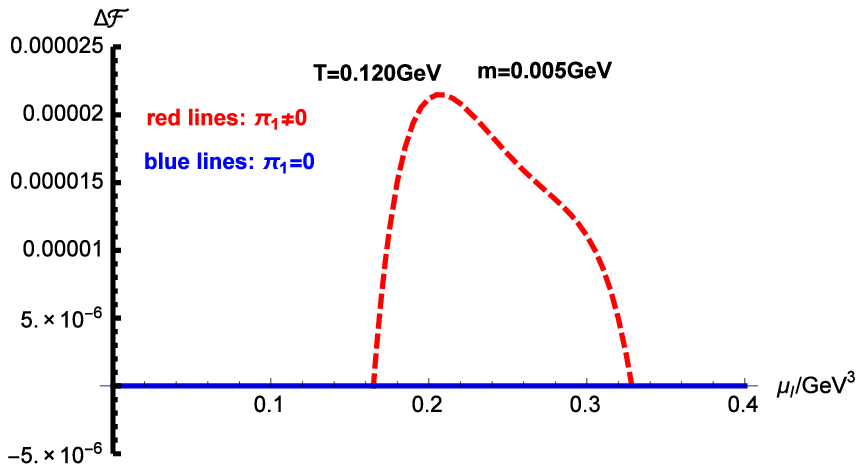}
\vskip -0.05cm \hskip 0.15 cm
\textbf{( a ) } \hskip 4.5 cm \textbf{( b )}  \hskip 4.5 cm \textbf{( c )} \\
\end{center}
\caption{Condensations $\pi_1$(a),$\sigma$(b) and free energy difference $\Delta \mathcal{F}$(c) as functions of isospin chemical potential $\mu_I$ when $T=0.120\rm{GeV}$.  In Panels.(a),(b) and (c), red and blue lines are solution from EoMs with and without pion condensation respectively.  Solutions with pion condensation exist only in the range $0.165\rm{GeV}<\mu_I<0.328\rm{GeV}$. From the free energy calculation in (c), solutions with pion condensation are thermodynamically unfavored.    }
\label{T120-cd}
\end{figure}

From our numerical analysis at high temperature, we find that the solutions with finite pion condensation exist until $\mu_I=0.127\rm{GeV}$. However, in the range $0.114\rm{GeV}<\mu_I<0.127\rm{GeV}$, all the solutions with pion condensation become thermodynamically unstable. To show this fact, we take $T=0.120\rm{GeV}$ as an example and plot the numerical results in Fig.\ref{T120-cd}.

From Fig.\ref{T120-cd}(a), we could see that solutions with pion condensation exist only in the range $0.165\rm{GeV}<\mu_I<0.328\rm{GeV}$, while non-trivial solutions without pion condensation appears in the whole range of $\mu_I$. In the solutions without pion condensation, $\sigma$ decreases with the increasing of $\mu_I$. From the free energy results in Fig.\ref{T120-cd}(c), we find that all the solutions with pion condensation have a larger free energy than solutions without pion condensation with the same $T,\mu_I$, which indicates no pion condensation in thermodynamical stable phase. As a result, no pion condensation exist above $T=0.114\rm{GeV}$.

\begin{figure}[h]
\begin{center}
\epsfxsize=7.5 cm \epsfysize=7.5 cm \epsfbox{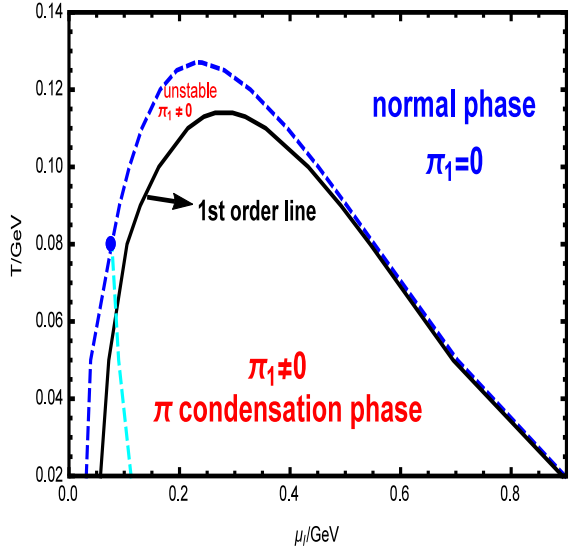}
\end{center}
\caption{Phase boundary of pion condensation in $\mu_I-T$ plane. Inside the region surrounded by the blue dashed line and the $\mu_I$ axis, there are solutions with non-zero pion condensation, while outside the region no solutions with pion condensation are founded. From free energy calculation, only the solutions inside the region surrounded by the black solid line and the $\mu_I$ axes are thermodynamically stable. The solutions in the region between the blue dashed line and black solid line are thermodynamically unstable. The black solid line is the phase boundary between the pion condensation phase and normal phase. It is also labeled that between the blue and cyan dashed lines there are triple solutions with non-zero pion condensation. The end point of this region is the blue dot, locating at $\mu_I\approx 0.077\rm{GeV},T\approx0.080\rm{GeV}.$ }
\label{phasediagram}
\end{figure}

Based on all of the above discussions, we summarize the results in Fig.\ref{phasediagram}. From the figure, the range of solutions with pion condensation is surrounded by the blue dashed line and the horizontal axes. Outside this range there would not be solutions with pion condensation.  However, the solutions with pion condensation inside this region are not always thermodynamically stable. Only the area surrounded by the black solid line in Fig.\ref{phasediagram} is related to the stable solutions. Also, the triple solutions area is separated out by blue, cyan dashed lines and the horizontal line. It is also observed that along the phase boundary there is a first order phase transition between pion condensation phase and normal phase.

\section{Conclusion and discussion}
\label{sum}

We investigate QCD phase transition, especially transition between pion condensation phase and normal phase, in a soft-wall AdS/QCD model. Introducing an interpolating dilaton filed as Refs.\cite{Chelabi:2015cwn,Chelabi:2015gpc,Li:2016gfn,Li:2016smq} and AdS-RN black hole solution with finite isospin chemical potential, we derive the equations of motion for the scalar, pseudo-scalar and axial vector filed. Then we calculate the near boundary expansion of those fields and match the integral constants to the quark mass, chiral condensation and pion condensation. The existence of horizon at finite isospin chemical potential makes the solution easy to diverge. Only at certain value of chiral and pion condensation the solution could be regular everywhere. As a result, requiring the regularity of physical solution, chiral and pion condensation could be solved self-consistently from the equations of motion. Using the 'shooting method', we solve the temperature and isospin chemical potential dependence of condensates when the quark mass $m=0.005\rm{GeV}$, around the physical value.

It is easy to check that there are two kinds of solutions. If one set the pion field and axial vector field to be zero, the solution would be reduced to those without pion condensation as discussed in \cite{Chelabi:2015cwn,Chelabi:2015gpc,Li:2016gfn,Li:2016smq}. The other kind is with finite pion condensation. The numerical results are summarized in Fig.\ref{phasediagram}. It shows that the pion condensation only exists in a bounded region surrounded by the black solid line and the $\mu_I$ axes in Fig.\ref{phasediagram}. When a system goes across the phase boundary, it will undergo a first order phase transition between pion condensation phase and normal phase without pion condensation.

In more details, the numerical calculation shows that the solution with pion condensation only exist below $T=0.127\rm{GeV}$. At a certain temperature below $T=0.127\rm{GeV}$, solutions with pion condensation only exist in a short range of $\mu_I$. The region for solutions with finite pion condensation are bounded by the blue dashed line and the $\mu_I$ axes. In this region, when $T<0.080\rm{GeV}$, there are triple solutions for a certain value of $\mu_I$. We labeled this region with the blue, the cyan dashed lines and the $\mu_I$ axes in Fig.\ref{phasediagram}. The triple solution structure strongly indicates a first order phase transition in small $\mu_I$, and the free energy calculation confirms it. At large $\mu_I$ with small temperature $T<0.080\rm{GeV}$, pion condensation decreases to zero continuously. This behavior looks like a second order phase transition, since the order parameter is continuous. However, from the free energy comparison, we find that it is actually a first order phase transition. The first order phase transition at large $\mu_I$ is in agreement with lattice simulations from \cite{Kogut:2002zg}. However, the lattice simulations\cite{Brandt:2017oyy,Kogut:2002zg} at small $\mu_I$ strongly indicates a second order phase transition in small $\mu_I$. This might be related to the probe limit approximation we used in this model. The background metric are simply taken as the AdS-RN black hole with isospin chemical potential, which could be seen as the first several expansion near the boundary $z=0$.  At large $\mu_I$ and $T$, the probe limit could be a good approximation, since at this range the horizon are very small. But at small $\mu_I$ and $T$, the horizon becomes large. Thus, significant corrections might be obtained from the full dynamics. We will leave this study to the future.

Further, when $0.080\rm{GeV}<T<0.114\rm{GeV}$, we find that the region of triple solutions disappears. In this temperature region, pion condensation is single valued at each $\mu_I$. At certain value of $\mu_I(T)$, pion condensation starts to grow up. Then after a short and rapid growth, pion condensation begin to decrease. This behavior is very similar to the lattice simulations at $\mu_I$ in \cite{Kogut:2002zg}(See Fig.\ref{T100-cd}(a) and Fig.\ref{lattice-phase}(b)). It looks like a second order phase transition from normal phase to pion condensation phase. However, the free energy calculation gives a first order phase transition again. Further confirmation from a full dynamical soft-wall model might be necessary to improve the small $\mu_I$ results. In the large $\mu_I$ part, qualitatively, it is almost the same as small $T$. Thus, it is coincident with the lattice simulation\cite{Kogut:2002zg}.

Then, in the temperature region $0.114\rm{GeV}<T<0.127\rm{GeV}$, though there are solutions with pion condensation, the free energy of these solutions are larger than that of solutions without pion condensation. In fact, these solutions are thermodynamically disfavored. Therefore, pion condensation can not survive above $T=0.127\rm{GeV}$.

Finally, it is also worth to mention that after the first order phase transition, the full condensation $\tilde{\sigma}$ jumps to a higher value, showing an enhancement effect of $\mu_I$ on chiral symmetry breaking at the transition point. This phenomena is in agreement with the study in hard-wall model \cite{Nishihara:2014nsa,Nishihara:2014nva}. But different from these previous studies, at even larger $\mu_I$, $\tilde{\sigma}$ will decrease with the increasing of $\mu_I$, which might need confirmation in a full dynamical model.

From the above discussions, when $\mu_I$ is not too large, the soft-wall model discussed in this work gives a first order phase transition from  normal phase to pion condensation phase with the increasing of $\mu_I$. The phase transition is accompanied with enhancement of chiral symmetry breaking.  Then when $\mu_I$ is sufficiently large, pion condensation will decrease and jump down to zero at a first order transition point at large $\mu_I$. After this transition, the system jumps to the normal phase and chiral symmetry would be restored. The large $\mu_I$ behavior of transition is in agreement with lattice simulation. However, the small $\mu_I$ transition needs to be check in a full dynamical holographic model, with the self-consistent interaction of gluo-dynamics and chiral dynamics.

\vskip 1.5cm
{\bf Acknowledgement}
\vskip 0.2cm

M.L. is supported by the National Natural Science Foundation of China(No.11505118). D.L. is supported by the National Natural Science Foundation of China(Nos.11805084 and 11647141) and the PhD Start-up Fund of Natural Science Foundation of Guangdong Province(No.2018030310457). S.H. is supported from Max-Planck fellowship in Germany and the German-Israeli Foundation for Scientific Research and Development.

\end{document}